\documentclass[prl,aps,groupedaddress,showpacs,twocolumn]{revtex4}
\usepackage{amssymb,amsmath,amsfonts,epsfig,graphicx,latexsym,bm,color,subfigure}

\begin{document}
\title{Light-induced coherence in an atom-cavity system}
\author{Christoph Georges$^{1}$, Jayson G. Cosme$^{1,2}$, Ludwig Mathey$^{1,2}$, and Andreas Hemmerich$^{1,2}$ \footnote{e-mail: hemmerich@physnet.uni-hamburg.de} }
\affiliation{$^{1}$ Institut f\"ur Laser-Physik and Zentrum f\"ur Optische Quantentechnologien, Universit\"at Hamburg, D-22761 Hamburg, Germany}
\affiliation{$^{2}$The Hamburg Center of Ultrafast Imaging, Luruper Chaussee 149, D-22761 Hamburg, Germany}
\date{\today}

\begin{abstract}
We demonstrate light-induced formation of coherence in a cold atomic gas system that utilizes the suppression of a competing density wave (DW) order. The condensed atoms are placed in an optical cavity and pumped by an external optical standing wave, which induces a long-range interaction mediated by photon scattering and a resulting DW order above a critical pump strength. We show that light-induced temporal modulation of the pump wave can suppress this DW order and restore coherence. This establishes a foundational principle of dynamical control of competing orders analogous to a hypothesized mechanism for light-induced superconductivity in high-$T_c$ cuprates. 
\end{abstract}

\bibliographystyle{prsty}

\maketitle
The simulation of complex electronic matter in a simplified optical lattice often requires tailor-made band structures \cite{Lew:07}. A celebrated method to this end is Floquet engineering, i.e., the application of modulation techniques, where typically the entire lattice is rapidly driven by light along a closed trajectory \cite{Eck:05, Lig:07, Zen:09, Str:11, Eck:17}. This renormalizes the tunneling amplitudes, which can become complex valued, such that a Peierls phase arises, similarly as for electrons in magnetic fields, and the resulting band structure can acquire a topologically nontrival character \cite{Str:11, Eck:17}. A common feature, however, is that the magnitudes of the dominant effective tunneling amplitudes generally decrease, and hence these modulation techniques typically lead to a decrease of coherence.

\begin{figure}
\includegraphics[scale=0.45, angle=0, origin=c]{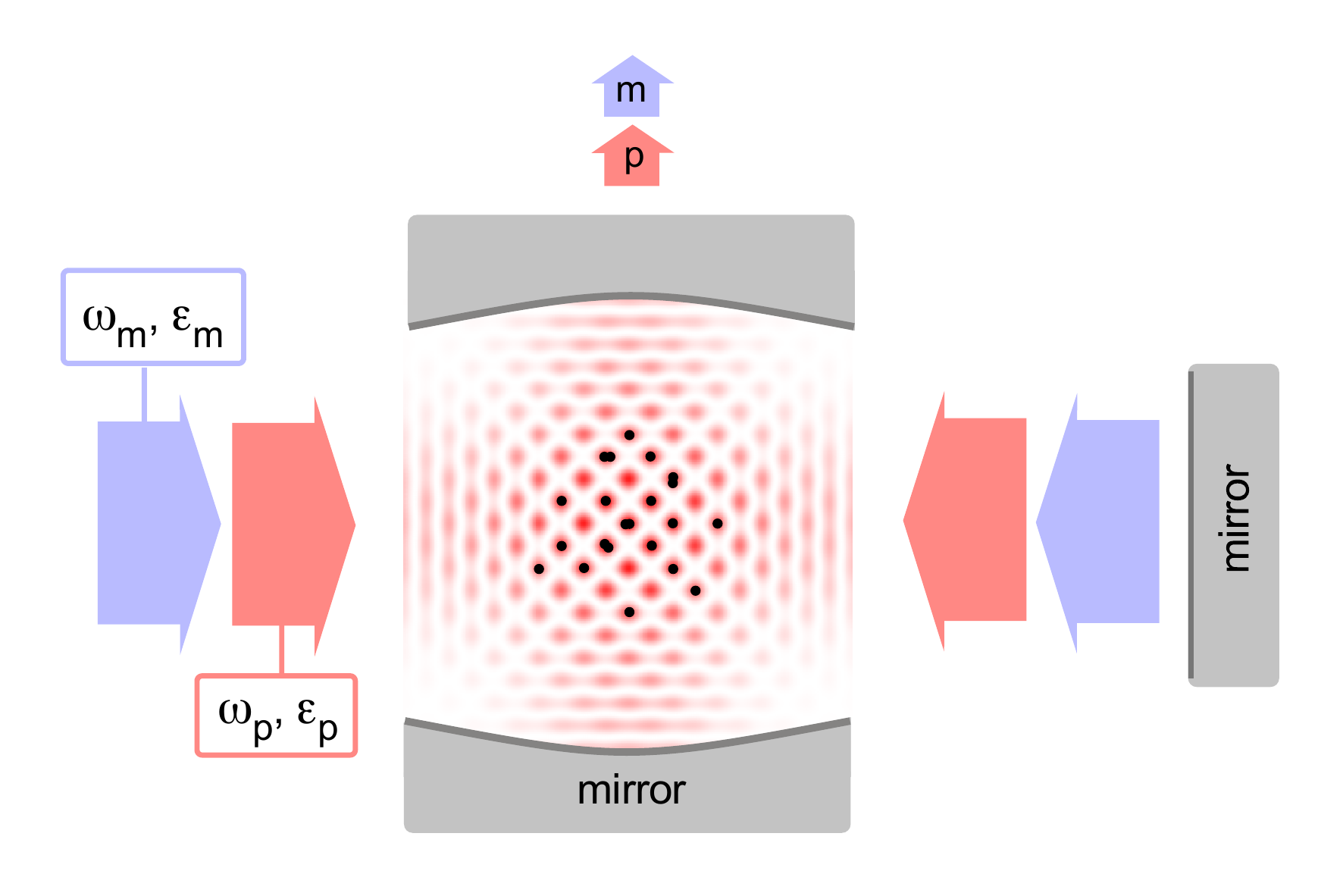}
\caption{\label{Fig.1} Bose-Einstein condensed rubidium atoms are held inside an optical cavity by a magnetic trap. An optical standing wave with frequency $\omega_p$ and strength $\varepsilon_p$ (red arrows) pumps the atoms. A second weaker optical standing wave at frequency $\omega_{m} = \omega_p + \delta_{m}$ and strength $\varepsilon_m$ (blue arrows) produces an amplitude modulation of the pump wave at the difference frequency $\delta_{m}/2\pi$.}
\label{fig:setup}
\end{figure}

Light-induced modulation techniques have also been applied to electronic condensed matter culminating in the spectacular observation that coherence can be increased. For example, superconductivity in high-$T_c$ cuprate compounds could be extended over short times well above $T_c$ of the nonmodulated systems and even up to room temperature \cite{Fau:11, Hu:14, Kai:14, Foe:14}. A generic example is La$_{1.8-x}$Eu$_{0.2}$Sr$_{x}$CuO$_4$ (LESCO$_x$), where due to the emergence of a competing 1D charge density wave (CDW) around the commensurate doping $x=1/8$ the critical temperature for superconductivity is significantly reduced \cite{Fau:11}. A proposed explanation is that around $x=1/8$ a Wigner type crystallization gives rise to a buckling of the CuO$_4$ octahedra, known as low temperature tetragonal phase. Using LESCO$_{1/8}$ it was shown that optical excitation in the near-infrared resonant with a CuO stretch mode could restore superconductivity at 10 K. The details of the underlying physical mechanisms are complex and still controversially debated and hence the identification of analogous physics in a more easily analyzed ultracold gas scenario, discussed in this Letter, appears highly welcome and instructive. 

\begin{figure*}[hbt]
\includegraphics[scale=0.45, angle=0, origin=c]{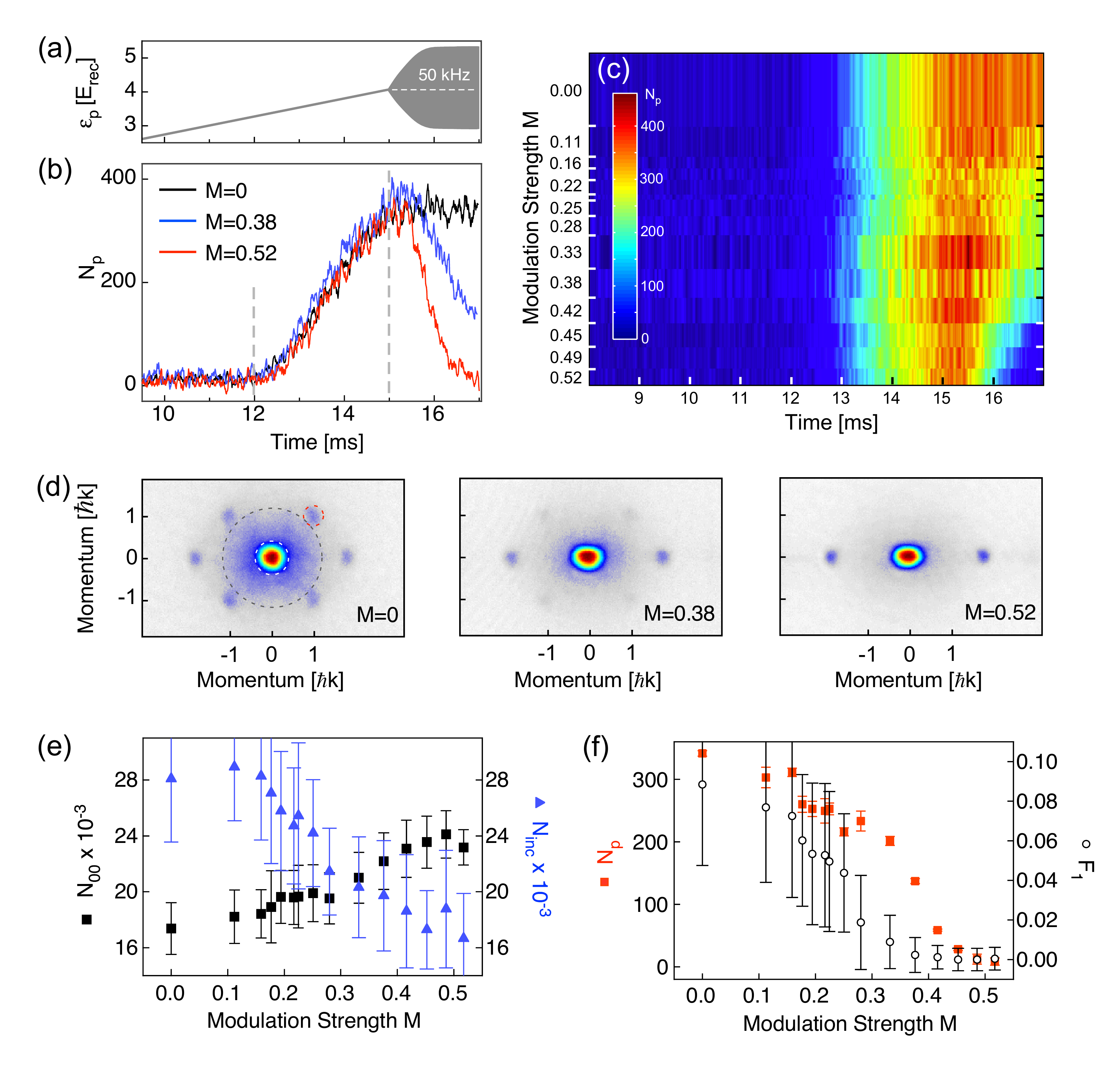}
\caption{\label{Fig.2} (a) Time sequence for pump and modulation beams. (b) The intracavity photon number $N_p$ is recorded versus time for three final values of the modulation strength $M$. (c) The intracavity photon number $N_p$ is plotted versus the final modulation strength $M$ and time, with the pump strength and the modulation strength ramped up according to (a). (d) Momentum spectra recorded after 2 ms of modulation according to the traces shown in (b). The dashed gray, white and red circles illustrate the locations of the incoherent fraction, the occupation of the BEC mode $(0,0) \hbar k$, and the atoms in the $(1,1) \hbar k$ momentum mode, respectively (see text). (e) The number of atoms $N_{00}$ in the BEC mode (black squares) and the number $N_{\textrm{inc}}$ of incoherent atoms. (f) The fraction $F_1$ of atoms in the first excited momentum modes at $(\pm 1,\pm 1)\hbar k$ (open circles) and the intracavity photon number (red squares) after 2 ms of modulation are plotted versus $M$. Because we do not make assumptions about possible systematic contributions, the error bars show the standard deviations for typically hundred repeated individual measurements. The number of atoms initially prepared in the BEC was $N_a = 6.5 \times 10^4$ and the modulation frequency was $\delta_{m}/2\pi = 50\,$kHz.}
\label{fig:data}
\end{figure*}

As suggested by recent theoretical work \cite{Cos:18}, the emergence of a competing CDW phase in LESCO$_x$ finds an analog for a Bose-Einstein condensate (BEC) placed inside a high finesse optical cavity and pumped sideways by an optical standing wave. In this system intracavity photons provide an infinite range retarded interaction among the atoms. At a critical pump strength a phase transition occurs from a spatially homogeneous BEC phase to an emergent particle density wave (DW) stabilized by an intracavity optical lattice \cite{Dom:02, Bla:03, Bau:10, Arn:12, Rit:13, Kli:15, Vai:18}. Close to the phase boundary the physics of this phase transition is captured by the Dicke model \cite{Dic:54, Hep:73}. The DW state is a superradiant Dicke state \cite{Gro:82}, for which the emergent Bragg grating of the atomic density scatters photons between the pump and the cavity mode. The condensate fraction of the atomic cloud drops sharply due to the onset of the competing density order. This phase transition displays a qualitative similarity to the competition of CDW order and superconductivity in LESCO, where BEC is the analogue of superconducting order, and each of these orders competes with a density order. In this Letter, confirming theoretical predictions in Ref.\cite{Cos:18}, we demonstrate that in analogy to light-induced superconductivity in LESCO the DW order can be suppressed by amplitude modulation of the pump wave such that the condensate fraction is significantly increased. We show that both mean-field (MF) and truncated Wigner (TW) simulations \cite{Bla:08, Pol:10} support our observation, as well as a high-frequency Magnus expansion that provides a renormalized low energy Hamiltonian, which allows for a physically consistent picture.

Our experimental setup is described in detail in Refs.\cite{Kes:14, Kes2:14, Kli:15}. An elongated BEC of $N_a \approx 10^5$ $\mathrm{^{87}Rb}$ atoms, prepared in the $|{F=2,m_F=2}\rangle$ state, is held in a magnetic trap. As illustrated in Fig.~\ref{fig:setup}, the atoms are pumped by an optical standing wave with frequency $\omega_p$ and wavelength $\lambda = 803\,$nm far detuned from the relevant atomic resonances, the atomic $D_{1}$ and $D_{2}$ lines at $795\,$nm and $780\,$nm, respectively. Its strength $\varepsilon_p$  is parametrized in terms of the induced antinode light shift in units of the recoil energy $E_{\textrm{rec}}\equiv \hbar^2 k^2/ 2m$ with $k = \omega_p/c$ and $m$ denoting the atomic mass. The field decay rate $\kappa = 2\pi \times 4.5\,$kHz of the optical cavity is smaller than $2\,\omega_{\mathrm{rec}} \equiv 2 E_{\textrm{rec}} / \hbar = 2\pi \times 7.1\,$kHz. The high finesse of the TEM$_{00}$ mode ($\mathcal{F} = 3.44 \times 10^5$) together with its narrow beam waist ($w_0 \approx 31.2\, \mu$m) yield a Purcell factor $\eta_\mathrm{c} \approx 44$ \cite{Pur:46, Tan:11}. Due to the details of the $D_{1,2}$ lines, the maximal coupling to the $|{F=2,m_F=2}\rangle$ atoms arises for left circularly polarized intracavity photons. For a uniform atomic sample, the TEM$_{00}$ resonance frequency for left circularly polarized light is dispersively shifted by an amount $\delta_{-} = \frac{1}{2} N_a \, \Delta_{-}$ with a light shift per photon $\Delta_{-} = \,- 2\pi \times 0.36\,$Hz. Hence, with $N_a = 6.5 \times 10^4$ atoms $\delta_{-} = - 2 \pi \times 12$~kHz, which amounts to $- 2.6\,\kappa$, i.e., the cavity operates in the regime of strong cooperative coupling. An additional laser beam with frequency $\omega_m$ and strength $\varepsilon_m$ is superimposed upon the pump beam as indicated in Fig.~\ref{fig:setup}. It induces an amplitude modulation of the pump wave without depleting its intensity, such that no trivial suppression of the DW order can arise. After BEC formation, the pump strength is ramped up until scattering into the cavity is observed and DW order is formed. Finally, the strength of the modulation beam $\varepsilon_m$ is ramped up. The light transmitted through the cavity is observed to determine the intracavity photon number. After a variable waiting time the atoms are rapidly released from the magnetic trap and their momentum spectrum is recorded by imaging the atomic cloud after 25 ms of ballistic expansion. For the observations shown here, we have chosen $\omega_p$ to be detuned from the effective cavity resonance frequency by $\delta_{\textrm{eff}}/2\pi = -38\,$kHz, since for this setting a stable DW phase can be formed. The modulation beam is detuned from the pump beam by $\delta_m/2\pi = 50\,$kHz. Over a wide range, the observations reported here do not sensitively depend on the exact value of $\delta_m$. Smaller $\delta_m$ are generally associated with increased heating and atom loss. At certain modulation frequencies, for example at twice the vibrational frequency of the pump lattice ($\delta_m/ 2 \pi \approx \pm 24\,$kHz), rapid resonant excitation occurs. 

In Fig.~\ref{fig:data} the observations are summarized. A typical experimental time sequence is illustrated in Fig.~\ref{fig:data}(a). Initially, the pump strength $\varepsilon_p$ is ramped up from zero to about $4\,E_{\mathrm{rec}}$ during 15 ms. At $t = 12\,$ms the critical pump strength is reached, where DW order is formed leading to the emergence of an intracavity lattice as observed in Fig.~\ref{fig:data}(b). After the final value of $\varepsilon_p$ is reached at $t = 15\,$ms the modulation beam is ramped up during 1 ms. As is seen in Fig.~\ref{fig:data}(b), depending on the modulation strength $M \equiv 2 \sqrt{\varepsilon_m \varepsilon_p}/(\varepsilon_m + \varepsilon_p)$ \cite{Mod}, the intracavity light field is partially or completely suppressed. A more complete picture is given in Fig.~\ref{fig:data}(c), where the intracavity photon number is plotted for 15 different values of $M$. In Fig.~\ref{fig:data}(d) momentum spectra are recorded after 2 ms of modulation according to the traces shown in Fig.~\ref{fig:data}(b). From such spectra we derive the populations of three subsamples of atoms: the amount $N_{00}$ of atoms condensed at zero momentum $(0,0) \hbar k$, approximately enclosed by the dashed white circle, the number of atoms in the incoherent fraction $N_{\textrm{inc}}$ approximately located between the gray and white dashed circles, and the number of atoms $N_{\nu\mu}$ populating the higher order modes at momenta $(\nu,\mu)\hbar k$ with $\nu,\mu = \pm 1$, and hence the fraction of atoms $F_{1} \equiv \Sigma_{\nu,\mu = \pm 1} N_{\nu\mu}/N_{00}$. To this end, the momentum spectra are fitted by a sum of a Gaussian, accounting for the incoherent fraction $N_{\textrm{inc}}$, and an inverted parabola according to the expected Thomas-Fermi density profile of the BEC mode occupation $N_{00}$. The populations of the $(\nu,\mu)\hbar k$ modes are obtained by summing over the densities within a spherical region (red dashed circle) surrounding the respective Bragg resonances. The emergence of an incoherent fraction indicates that DW order is associated with a significantly reduced degree of coherence as compared to BEC order, which may partly be attributed to collisional dephasing in the self-organized lattice. As $M$ is increased, the momentum spectra approach those of a pure BEC merely subject to the pump wave, with $N_{\textrm{inc}}$ and $F_{1}$ becoming increasingly small. In Fig.~\ref{fig:data}(e) $N_{\textrm{inc}}$ (blue triangles) and $N_{00}$ (black squares) are plotted versus $M$. It is seen that as $M$ increases, $N_{00}$ increases by about $30 \%$, while $N_{\textrm{inc}}$ decreases by about $43 \%$. Due to heating and losses introduced with increasing modulation strength, the sum, which, before the modulation sets in, amounts to $N_{00} + N_{\textrm{inc}} \approx 47.000$ drops by about $13\%$ for the maximally applied value of the modulation strength $M=0.52$. Furthermore, the fraction of atoms $F_{1}$ is nearly eliminated, as is shown in Fig.~\ref{fig:data}(f) (open circles). This indicates the disappearance of the density grating, which is also reflected by the vanishing of the intracavity photon number (red squares), also observed after 2 ms of modulation. Remarkably, not only $F_{1}$ is suppressed by the modulation but also the incoherent fraction of atoms $N_{\textrm{inc}}$ is significantly reduced into the BEC mode. 

\begin{figure}
\includegraphics[scale=0.33, angle=0, origin=c]{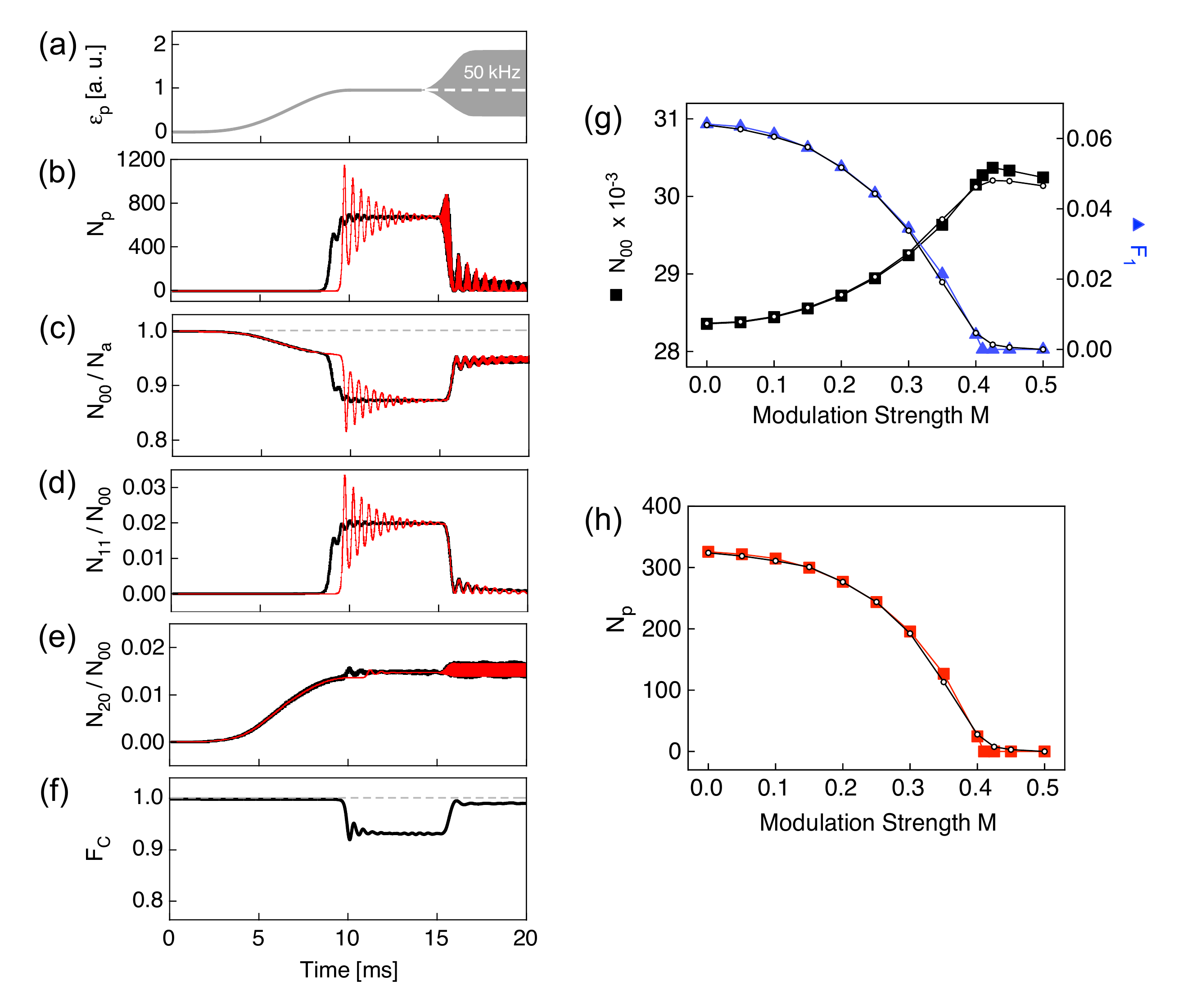}
\caption{\label{Fig.2} (a) Time sequence for pump and modulation beams with $\delta_{\textrm{eff}}/2\pi = -38\,$kHz and $\delta_{m}/2\pi = 50\,$kHz. In (b)-(e) MF (red traces) and TW calculations (black traces) are superimposed. (b) shows the intracavity photon number $N_p$. (c)-(e) show the corresponding relative populations of the zero momentum BEC mode (c), the first excited $(1,1) \hbar k$ mode (d), and the $(2,0) \hbar k$ mode (e). (f) shows the condensate fraction $F_c$ defined as the largest eigenvalue of the single-particle density matrix (see text). For all plots (a) - (f) the modulation strength is $M = 0.4$. (g) and (h), MF calculations of the population of the BEC mode (black squares in (g)), the fraction $F_1$ of atoms in the $(\pm 1, \pm 1) \hbar k$-modes (blue triangles in (f)), and the intracavity photon number $N_p$ (red squares in (h)) after 10 ms of modulation. The white disks compare calculations based on a TW approximation.}
\label{fig:mean_field}
\end{figure}

We analyze our findings by first employing a MF simulation based on Eq.(2) of the Supplemental Material of Ref.~\cite{Kli:15} and then a TW simulation \cite{Cos:18}. For the TW simulation we use an open system formalism that extends these MF equations by including the noise due to photon loss at the rate $\kappa$. While these equations of motion neglect low-momentum modes or $s$-wave scattering between atoms, we show that the basic phenomenon is reproduced in Fig.~\ref{fig:mean_field}. In Fig.~\ref{fig:mean_field}(a) the chosen time sequence is shown. In contrast to the experiment, instead of a linear ramp a smoother increase of $\varepsilon_p$ is chosen during the first 10 ms followed by a 5 ms long period during which $\varepsilon_p$ is kept constant, to give the system time to assume its equilibrium. As observed in the red traces in Figs.~\ref{fig:mean_field}(b)-\ref{fig:mean_field}(e), at the onset of DW order the mean field calculations show oscillatory dynamics, which is not seen in the experimental data. These oscillations are significantly reduced, if quantum fluctuations are included by using a TW approximation, as shown by the superimposed black traces. Further damping may be expected from binary collision dynamics not accounted for in the calculations \cite{Kon:14}. Figures.~\ref{fig:mean_field}(b)-\ref{fig:mean_field}(e) show that DW order, as in the experimental observations, is associated with the formation of an intracavity light field [Figures.~\ref{fig:mean_field}(b)], a depletion of the BEC mode [Figures.~\ref{fig:mean_field}(c)] and an excitation of the momentum modes $(\pm 1,\pm 1)\hbar k$ [Figures.~\ref{fig:mean_field}(d)]. The $(2,0)\hbar k$ momentum mode, shown in Figures.~\ref{fig:mean_field}(e), corresponds to an excitation along the pump wave and hence it becomes smoothly populated as the pump strength increases irrespective of the presence of DW order. When the modulation beam is activated at 15 ms, the DW order is suppressed and the BEC depletion is removed in agreement with the observations. Note that the intracavity field, if not entirely suppressed, oscillates with the modulation frequency $\delta_{m}$. The BEC mode population does not come entirely back to its original value. Rather a small oscillation of population between the BEC mode and the $(2,0)\hbar k$-mode arises at the frequency $\delta_{m}$. This is not surprising since even if the intracavity lattice is fully suppressed, the pump lattice continues to oscillate at the frequency $\delta_{m}$.

In Figures.~\ref{fig:mean_field}(g) and \ref{fig:mean_field}(h), we show MF and TW calculations associated with the observation in Figs.~\ref{fig:data}(e) and \ref{fig:data}(f), respectively. In Fig.~\ref{fig:mean_field}(g), we show MF calculations of the population of the BEC mode (black squares) and the fraction of atoms in the $(\pm 1, \pm 1) \hbar k$ modes (blue triangles) obtained after 10 ms of modulation, i.e. well after the system has relaxed to equilibrium. In order to account for the fluctuations associated with the damping of the cavity and to more realistically model the initial state by including quantum fluctuations in the BEC mode and vacuum fluctuations in the higher momentum modes and the cavity mode, we have also added calculations based on a TW approximation, shown by the white disks on top of both graphs. Fig.~\ref{fig:mean_field}(h) shows a MF calculation (red squares) and a TW calculation (white disks) of the intracavity photon number. Apart from transient behavior at the onset of DW order, as in Fig.~\ref{fig:mean_field}(b)-(e), MF and TW calculations only show minor deviations for the population of the different modes. In particular, the BEC mode shows an increase of population both for the MF and the TW simulation. The calculations shown in Fig.~\ref{fig:mean_field}(g),(h) qualitatively reproduce the increase of the population of the BEC mode $N_{00}$, the reduction of the population of the $(\pm 1, \pm 1) \hbar k$ modes, and the associated suppression of the intracavity field, observed according to Figs.~\ref{fig:data}(e) and (f). 

In order to explore the evolution of the coherence properties of the BEC, before the DW phase is entered, in the DW phase, and after modulation with $M=0.4$, we have plotted the condensate fraction $F_c$ in Fig.~\ref{fig:mean_field}(f). This quantity is directly related to the Penrose-Onsager criterion for condensates and is calculated as the largest eigenvalue of the single-particle density matrix, which quantifies the equal-time correlations between the momentum modes \cite{Lod:17}. As is seen in (f), $F_c$ is practically unity, without DW order, it is reduced by $7 \%$ as the DW phase is entered, and recovers to $99 \%$ after modulation is applied. The notable increase of the incoherent fraction of atoms $N_{\textrm{inc}}$ observed in the DW phase accounts for this coherence loss as well as the additional even larger coherence loss resulting from collision-induced fragmentation, not included in the present calculations.
 
An improved quantitative agreement should be possible, if collision-induced contact interaction is accounted for, which plays a significant role in our experiments due to condensate densities above $10^{14}\,$cm$^{-3}$ resulting from the necessity to confine the atoms well within the narrow cavity mode. Such calculations are expected to be numerically far more expensive and require further efforts beyond the scope of this work. Finally, we note, that a high-frequency Magnus expansion \cite{Buk:15, Zhu:16, Hem:10, Gol:14, Eck:15} shows that the effect of modulation can be understood in terms of a renormalization of the effective detuning $\delta_{\textrm{eff}}$ and the dc pump strength $\varepsilon_p$ such that the former is increased and the latter is decreased \cite{Cos:18}. Note that, different from renormalized tunneling in conventional lattice shaking, the renormalization here concerns terms of the Hamiltonian with many-body character. The recoil resolution of our cavity and the associated retardation of the cavity-mediated interaction is an essential feature of our system, which exhibits a many-body character beyond a parametrically driven Dicke model \cite{Bas:12, Chi:15}. 

\begin{acknowledgments}
This work was partially supported by DFG-SFB 925 and the Hamburg Center for Ultrafast Imaging (CUI). We thank Cristiane Morais Smith, Andrea Cavalleri, and Claus Zimmermann for useful discussions.
\end{acknowledgments}


\begin{thebibliography}{25}
\bibitem{Mod}
The modulation strength $M$ is defined as the interference contrast between the pump and the modulation beams with field amplitudes $\sqrt{\varepsilon_p}$ and $\sqrt{\varepsilon_m}$. Hence, $M$ takes its maximal value $M=1$ at $\varepsilon_m = \varepsilon_p$ and $M=0.5$ arises for $\varepsilon_m / \varepsilon_p = 7 - 4\sqrt{3} \approx 0.072$.
\end{thebibliography}
\end{document}